\documentclass[apj]{emulateapj}
\usepackage{epstopdf}
\def\gsim{\ \raise 3pt \hbox{$\rangle$} \kern -8.5pt \raise -2pt \hbox{$\sim$}\ }
\usepackage{graphicx,natbib,color,amsmath,subfigure}
\usepackage{ulem}
\newcommand{\blank}[1]{}
\usepackage{color}
\newcommand{\gfR}{} 

\newcommand{\dg}{} 
\newcommand{\gf}{} 
\newcommand{\gff}{\bf \color{red}}

\def\mw{{microwave}}

\begin{document}
\title{Energy Partitions and Evolution in a Purely Thermal Solar Flare
}
\author{Gregory D. Fleishman\altaffilmark{1,2}, Gelu M. Nita\altaffilmark{1}, Dale E. Gary\altaffilmark{1}}
\altaffiltext{1}{Center For Solar-Terrestrial Research, New Jersey Institute of Technology, Newark, NJ 07102}
\altaffiltext{2}{Ioffe Physico-Technical Institute, St. Petersburg 194021, Russia}

\begin{abstract}
This paper presents 
a solely thermal flare, which we detected in the microwave range from the thermal gyro- and free-free emission it produced. An advantage of analyzing thermal gyro emission is its unique ability to precisely yield the magnetic field in the radiating volume. When combined with observationally-deduced plasma density and temperature, these magnetic field measurements offer a straightforward way of tracking evolution of the magnetic and thermal energies in the flare. For the event described here, the magnetic energy density in the radio-emitting volume declines over the flare rise phase, then stays roughly constant during the extended peak phase, but recovers to the original level over the decay phase. At the stage where the magnetic energy density decreases,  the thermal energy density increases; however, this increase is insufficient, by roughly an order of magnitude, to compensate for the magnetic energy decrease.
When the magnetic energy release is over, the source parameters come back to nearly their original values. We discuss possible scenarios to explain this behavior.

\end{abstract}


\section{Introduction}

It is well established that to produce a flare requires magnetic free energy stored in the form of nonpotential coronal magnetic field (supported by coronal electric currents), which is somehow released and converted into other forms of energy---kinetic, thermal, and nonthermal. This process of magnetic energy release is commonly referred to as magnetic reconnection.

Flaring energy release is necessarily an extended process. Indeed, to supply a large flare with an energy of $\sim10^{32}$~erg at the expense of magnetic energy dissipation, one needs a rather large volume. For example, $\sim10^{32}$~erg is roughly the energy contained in a $100$~G magnetic field in a volume $\sim 3\cdot10^{29}$~cm$^3$. Since only free (nonpotential) magnetic energy is available for release, {the energy release region must occupy an even larger volume.}

If the magnetic reconnection/energy release occurs in a given limited volume, it is reasonable to anticipate that the magnetic field strength, energy density, and the total energy all decrease during the energy release stage. Thus, to probe the very process of energy release and transformation to other forms, it would be highly desirable to track all the energy components in the flaring volume. This is, however, a highly complicated task \citep{Emslie_etal_2012}. Indeed, although the thermal energy could often be reliably estimated with the soft X-ray (SXR) emission and some other means, the other energy components are very difficult to precisely measure at the flare energy-release site.

The thermal flare energy can often be understood as the plasma response to accelerated particle input; the relationship commonly referred to as the Neupert effect \citep{1968ApJ...153L..59N}. However, there are cases (cold flares) when no detectable heating is observed in spite of highly efficient acceleration. In particular, in the 30 Jul 2002 cold flare we studied recently \citep{Fl_etal_2011} we found that almost all available electrons were accelerated. It was not possible, however, to construct the complete energy balance, because the fast electrons deposited their energy into the chromosphere, as confirmed by footpoint hard X-ray (HXR) emission, although no significant evaporation occurred, and diagnostics of the chromospheric response could not be measured. 
From the gyrosynchrotron (GS) spectrum we estimated the magnetic field in the particle acceleration region, but obtained a magnetic field value that did not show any significant evolution. This can mean that, because the nonthermal electrons leave the acceleration region relatively quickly, and thus do not contribute to the total pressure in the flaring volume, the magnetic field is supported by {an} external pressure at roughly {its} original level.

On the other hand, in many events there is a preflare phase that is almost purely thermal \citep{Battaglia_etal_2009}, although showing a weak nonthermal component in many cases \citep{Asai_etal_2006, Asai_etal_2009, Altyntsev_etal_2012}. Since the nonthermal energy content is only minor over these preflare heating episodes, the nonthermal particles alone cannot account for the heating. Thus, the plasma must be somehow heated directly by the magnetic energy release, or indirectly, e.g., via dissipation of the turbulent motions and/or waves generated, perhaps, as a result of the magnetic energy release. But this implies that there can be cases when the nonthermal component is essentially nonexistent over the entire event duration, which would be
{manifested} as {a} purely thermal event.

Indeed, such cases have been observed in both microwave and X-ray domains. In particular, \citet{Gary_Hurford_1989} reported a simple microwave burst showing the spectral shape consistent with being produced by a purely thermal plasma (with a Maxwellian distribution without any nonthermal tail) over the entire evolution of the flare. From the sequence of spectral fits they evaluated the characteristic magnetic field at the radio source to be roughly constant at the level of $\sim770$~G.
Recently, \citet{liu_etal_2013} reported a thermal flare that occurred in a low-lying coronal loop. They derived evolution of the plasma temperature, but no direct data on the magnetic field at the source was available {due to lack of radio observations}.

\begin{figure} \centering
\includegraphics[width=0.32\columnwidth,clip]{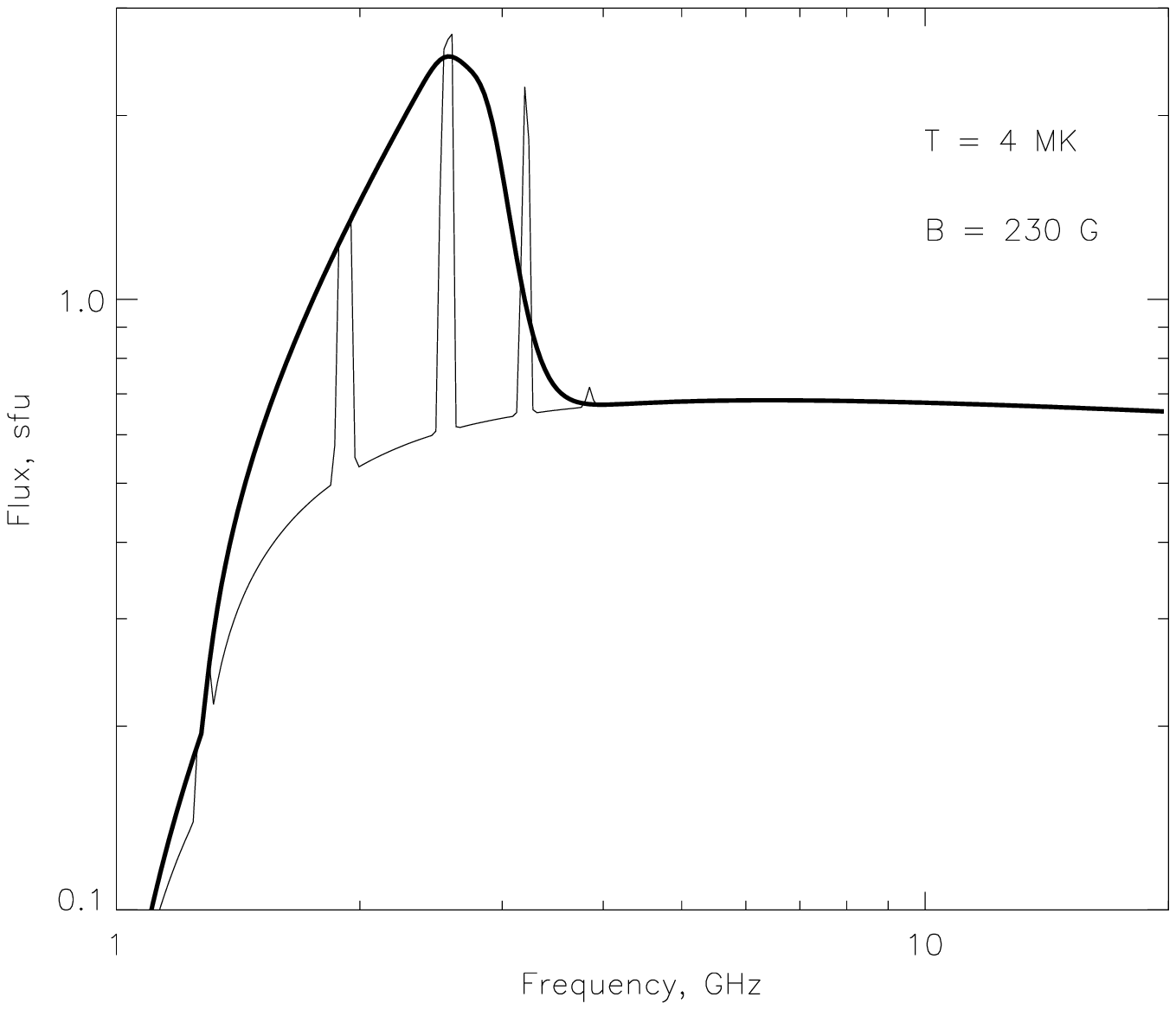}
\includegraphics[width=0.32\columnwidth,clip]{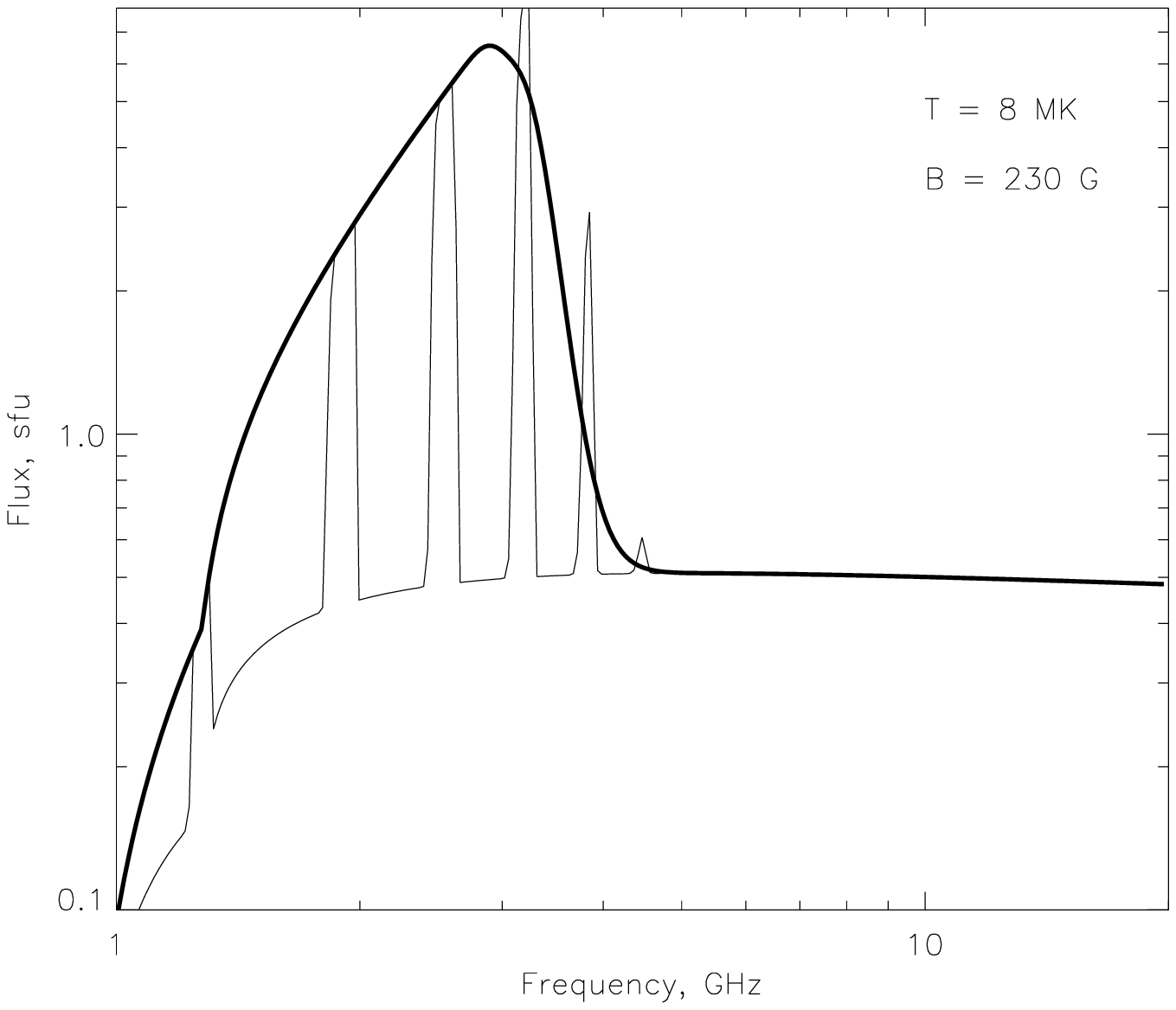}
\includegraphics[width=0.32\columnwidth,clip]{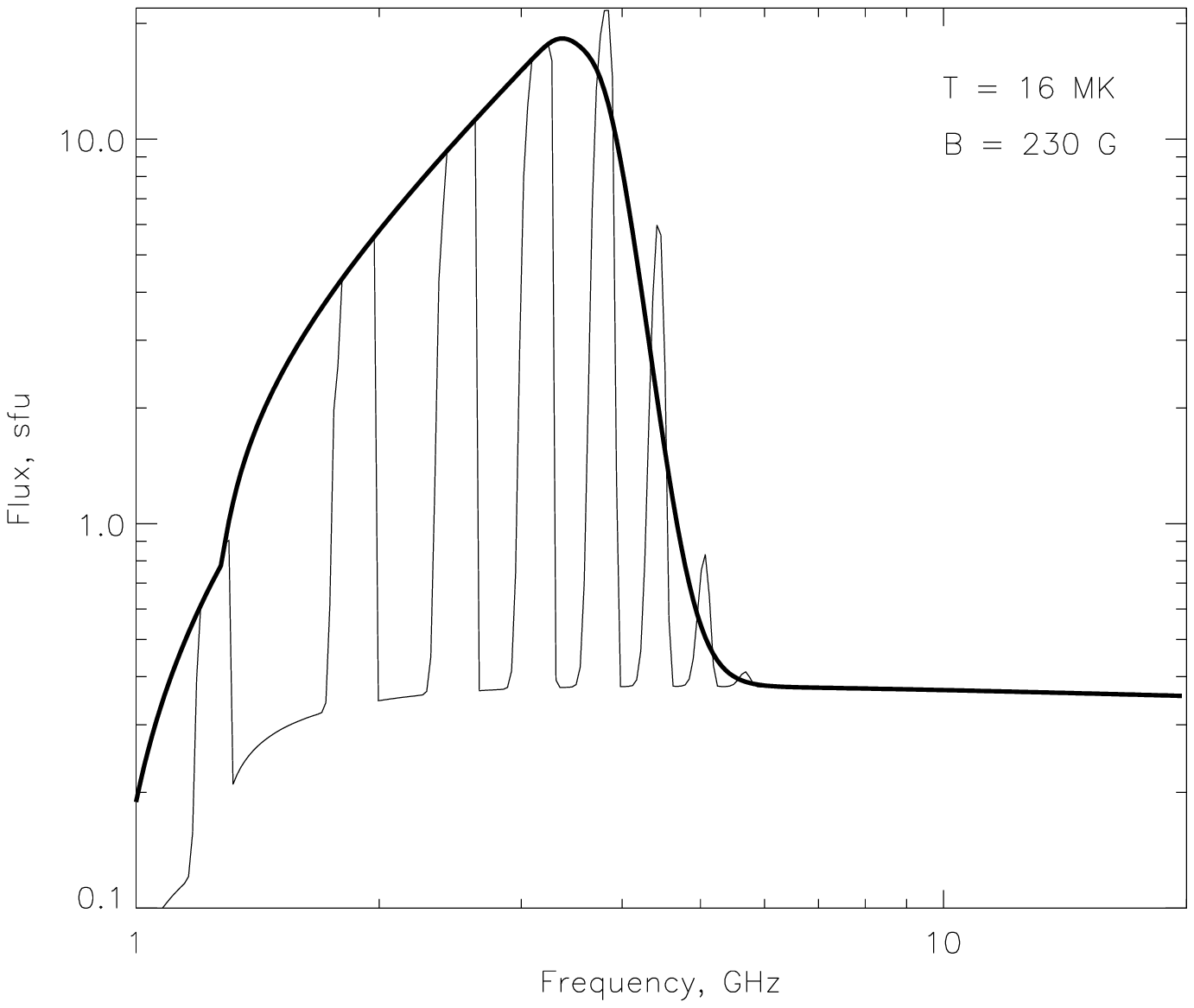}
\includegraphics[width=0.32\columnwidth,clip]{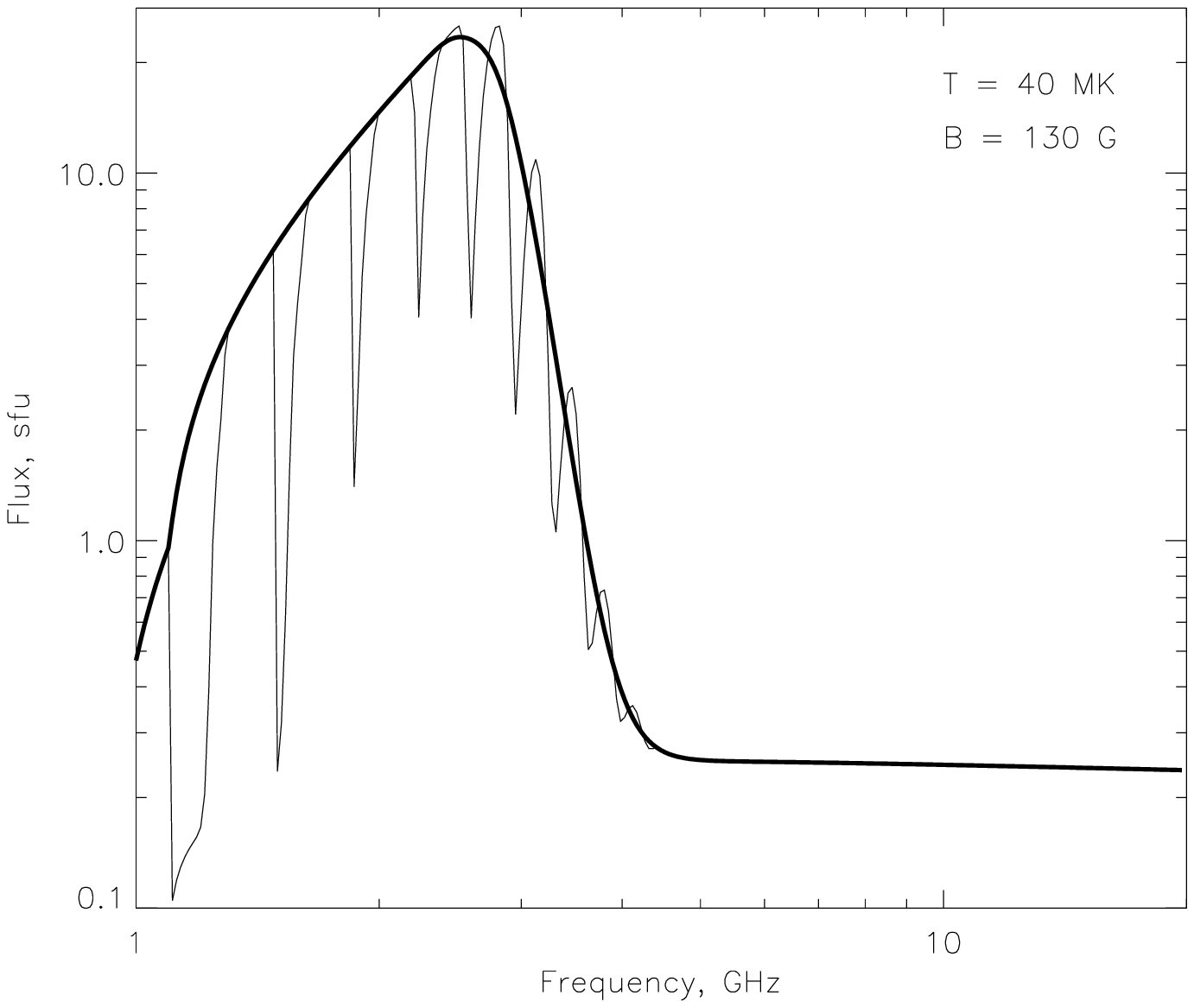}
\includegraphics[width=0.32\columnwidth,clip]{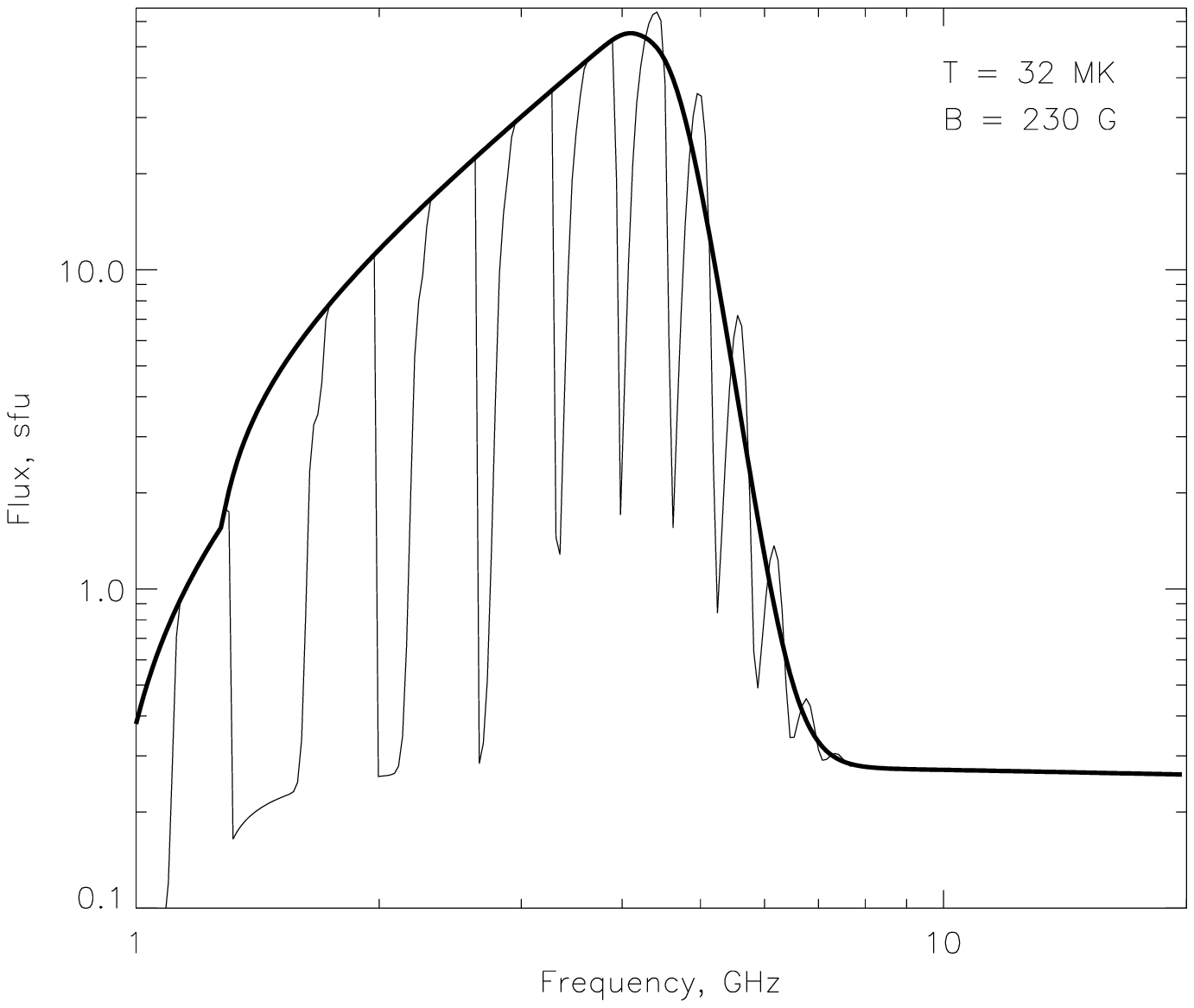}
\includegraphics[width=0.32\columnwidth,clip]{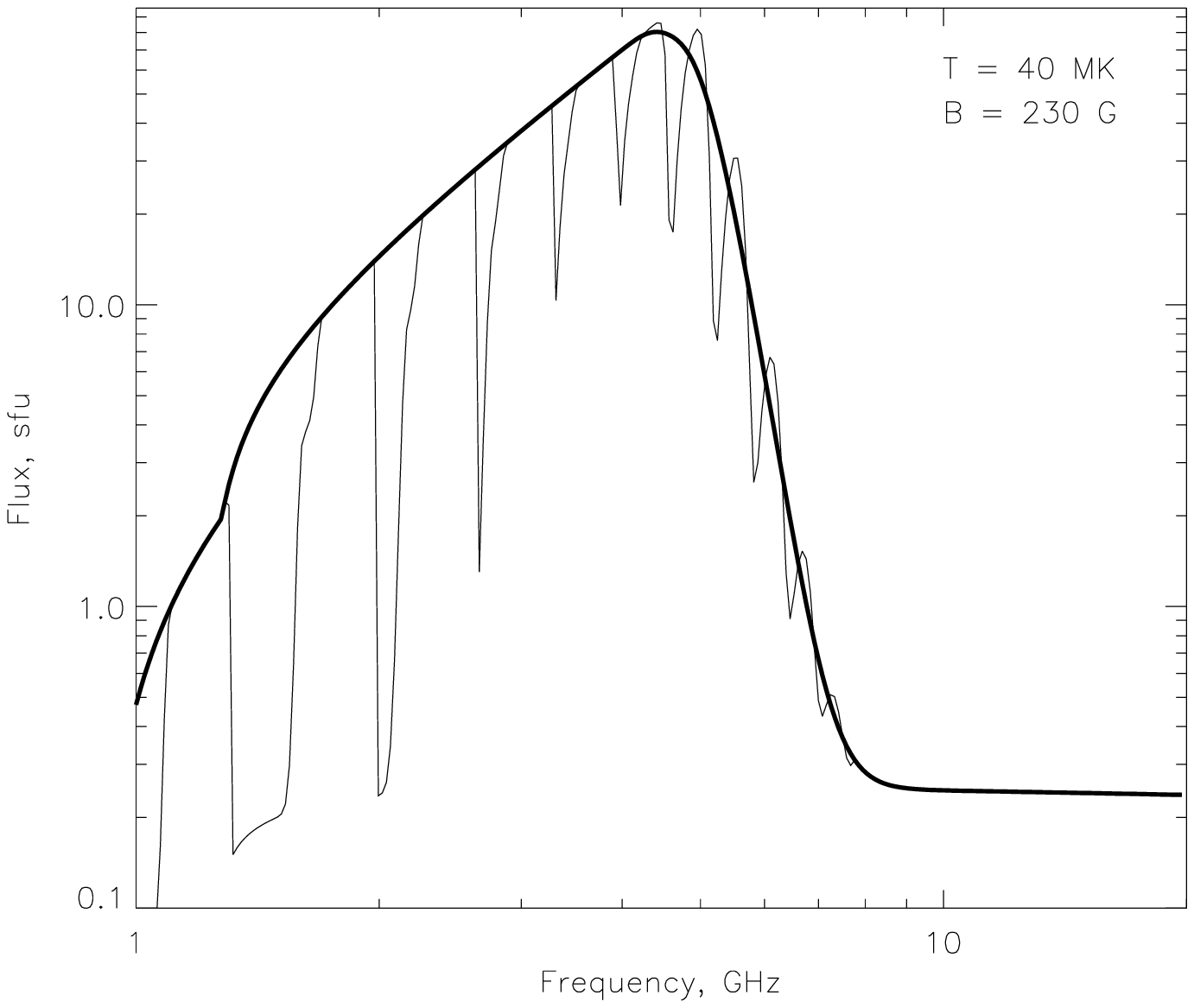}
\caption{\label{Fig_spectrum_example} Model GS + free-free spectra from a uniform plasma volume computed with an exact expressions \citep[specifically, the hybrid fast GS code,][thin curves]{Fl_Kuzn_2010} and continuous approximation (thick curves). The adopted parameters are: area $A=1600''^2$, depth $d=10''$, plasma density $n_0 = 10^{10}$~cm$^{-3}$, and viewing angle $\theta=75^\circ$. Values of the plasma temperature $T$ and magnetic field $B$ are shown in the panels. Note, that for a given magnetic field, the spectrum peak increases when the temperature increases. Thus, having a constant spectral peak while the plasma heating requires a corresponding decrease of the magnetic filed.
}
\end{figure}

Microwave data offer diagnostics of both magnetic field and thermal plasma.  For a purely thermal flare the microwave emission is produced by a combination of the thermal GS plus free-free emissions.
As an example, Figure~\ref{Fig_spectrum_example} displays  microwave spectra from a given volume uniformly filled with a hot plasma with various temperatures and having uniform magnetic field. The {exact} spectr{a (thin lines)} consist of a number of peaks corresponding to gyroharmonics (integer multiples of the gyrofrequency), whose flux level increases with frequency at low frequencies, where the emission at each gyroharmonic is optically thick and the corresponding brightness temperature is just the plasma kinetic temperature; thus the average flux level rises as $\propto f^2$. Then, above a certain frequency, which increases as the temperature and/or magnetic field increases, the emission becomes optically thin and so the flux level decreases quickly; thus, at even higher frequencies the flux level is determined by the optically thin free-free emission, which yields a flat spectrum at high frequencies.
{I}solated gyroharmonic emission has
been reported {for non-flaring active regions} \citep{Lang_etal_1987, Bogod_etal_2000}, and
from solar flares {\citep{Benka_Holman_1992}, but such reports are rare} because the magnetic field is spatially nonuniform at the radio source. Even a relatively minor nonuniformity of the magnetic field will smooth the isolated harmonics{, thus the thin line curves in Figure~\ref{Fig_spectrum_example} are essentially unobservable in most cases.  Instead,}
a continuum thermal spectrum {is observed (thick lines in Figure~\ref{Fig_spectrum_example}),} with flux density $\propto f^2$ at low frequencies, where the emission is optically thick, and a rapid (exponent-like) drop-off at higher frequencies, where the emission is optically thin; the falling part of the gyro-spectrum then gives way to a flat free-free spectrum.

The spectral peak of the thermal GS emission allows a precise measurement of the maximum magnetic field at the source, while the level of the optically thick part of the spectrum depends only on the product of the plasma temperature and source area. Above the spectral peak the microwave spectrum is flat due to the optically thin free-free emission at the level defined by the plasma density and temperature. Thus, microwave observations of a purely thermal source offer a straightforward way of estimating physical parameters needed to study the energy balance and evolution in such events. \cite{Altyntsev_etal_2012} have demonstrated the ability of thermal GS emission to yield magnetic field diagnostics at the energy release site. However, they used microwave data from a few discrete frequencies provided by the Nobeyama Radio Polarimeters (NoRP), RSTN, and Siberian Solar Radio Telescope (SSRT). As a result, the microwave spectral peak could not be precisely determined, which is needed for precise magnetic field diagnostics.

Here we use spectrally resolved Owens Valley Solar Array (OVSA) observations that provide us with much better measurements of the spectr{al} shape sufficient to firmly constrain the source parameters such as temperature by area product and the magnetic field.  We also use these exceptionally clean spectra as an opportunity to investigate the potential influence of a slight non-Maxwellian distribution, in the form of the kappa distribution. 

\section{Observations}

\begin{figure} \centering
\includegraphics[width=0.96\columnwidth,clip]{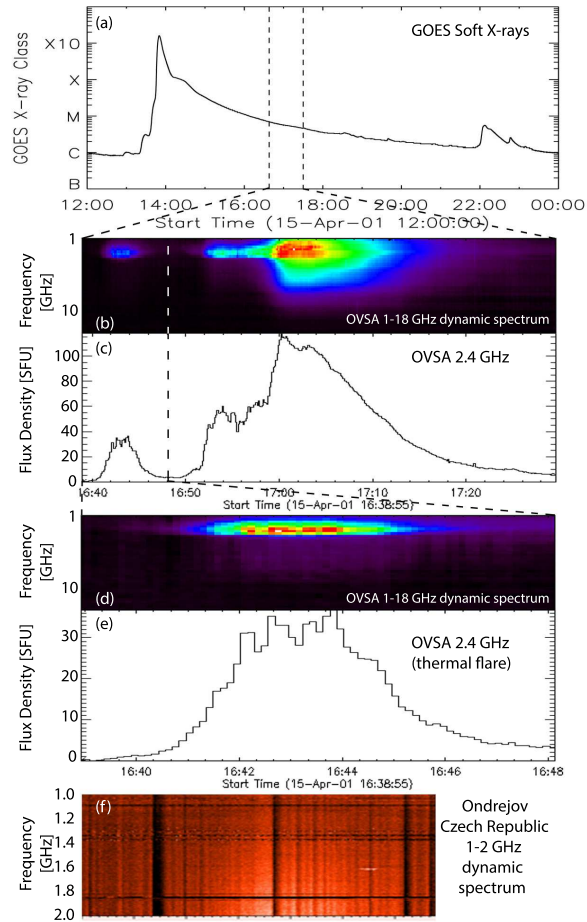}
\caption{\label{fig_OVSA_dynspec}The thermal flare event in the context of other radio activity. (a) The GOES (1-8 \AA) soft X-ray light curve spanning 12 hours (1200-2400~UT) on 2001 Apr 15, showing the large (X14.4) flare peaking around 13:50~UT and its long decay. The vertical dashed lines show the timerange shown in b-c.  (b) The OVSA dynamic spectrum from 1-18~GHz from 1640-1730~UT, showing the narrowband thermal flare (left of the white dashed line) and an increasingly nonthermal preflare phase leading to a larger event peaking at 1700 UT.  (c) The 2.4~GHz lightcurve corresponding to b.  To the left of the vertical dashed line is the timerange of the thermal flare shown in d-e. (d) The OVSA dynamic spectrum from 1-18~GHz from 1638-1648~UT, showing the narrow-band thermal flare, visible only in the range 1.0-3.2~GHz. (e) The 2.4~GHz lightcurve corresponding to d. (f) The 1-2~GHz dynamic spectrum from Ondrejov on the same timescale as in d-e.  The horizontal and vertical lines are instrumental artifacts.
}
\end{figure}

The period of thermal emission that we study manifested during the decay phase of one of the largest flares of solar cycle 23, the GOES soft X-ray class X14.4 event of 2001 Apr 15 that peaked $\sim$~13:45~UT (Figure~\ref{fig_OVSA_dynspec}a), and was spatially associated with the post-flare loops of that event.  However, we believe the thermal event should more-properly be considered as a 'precursor'  
of a much smaller event that peaked $\sim$~17:00~UT (Figure~\ref{fig_OVSA_dynspec}b-c).  Figure~\ref{fig_OVSA_dynspec}b shows the Stokes $I$ total power dynamic spectrum at 16:38-17:30 UT from the Owens Valley Solar Array \citep[OVSA;][]{ovsa_1984, Gary_Hurford_1994}, obtained from 40 frequencies roughly logarithmically spaced over the range 1-18~GHz.  To aid in visibility of the emission, the corresponding 2.4~GHz Stokes I total power lightcurve is shown in Figure~\ref{fig_OVSA_dynspec}c.  No significant circular polarization was seen during the event. The spectral shape of the emission, as we will show shortly, is consistent with purely thermal gyrosynchrotron emission during the small, isolated peak to the left of the dashed line in Figure~\ref{fig_OVSA_dynspec}b-c (16:40-16:48 UT), then it becomes increasingly nonthermal during the time to the right of the dashed line, leading up to the event peaking at 17:00~UT. Figure~\ref{fig_OVSA_dynspec}d-e show the OVSA dynamic spectrum and 2.4~GHz lightcurve for the period of the thermal flare, from 16:38-16:48~UT.  The narrow-band nature of the emission in Figure~\ref{fig_OVSA_dynspec}d is striking.  One sometimes sees highly nonthermal, coherent plasma emission in the decimeter range below 3~GHz \citep{Nita_etal_2004}, 
which OVSA's frequency and time resolution may not resolve.  However, Figure~\ref{fig_OVSA_dynspec}f shows much higher frequency- and time-resolution data from the Ondrejov observatory \citep{ondrejov}, on the same time scale as Figure~\ref{fig_OVSA_dynspec}d-e.  The Ondrejov data confirm the smooth continuum nature of the emission, consistent with our interpretation of thermal gyrosynchrotron emission, and eliminates the possibility that the radio spectrum is affected by coherent plasma emission, which would be a signature of nonthermal particles.

\subsection{Context Observations}
\begin{figure} \centering
\includegraphics[width=0.96\columnwidth]{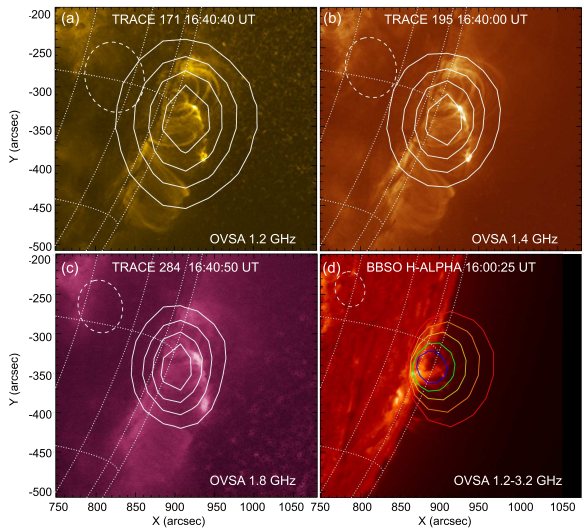}\\
\caption{\label{fig_OVSA_imaging} Radio images and other context imaging data. (a) 1.2~GHz OVSA (contours) on top of TRACE 171~\AA\ (image), (b) 1.4~GHz OVSA (contours) on top of TRACE 195~\AA\ (image), (c) 1.8~GHz OVSA (contours) on top of TRACE 284~\AA\ (image), and (d) OVSA 50\% contour at each of 1.2 (red), 1.4 (orange), 1.8 (yellow), 2.4 (green) and 3.2~GHz (blue), on top of BBSO H$_\alpha$ (image), showing a slight dispersion in position with frequency. {See on-line movies showing the source variation either with frequency or with time.}
}
\end{figure}

The Transition Region And Coronal Explorer \citep[TRACE;][]{trace} and Solar Heliospheric Observatory \citep[SoHO;][]{soho,soho_eit} targeted this AR and revealed a system of postflare coronal loops from the X14.4 event. The radio burst originated from the same area, but it is not possible to associate the radio source with a given single EUV loop, and in fact, as we will see, the radio source is far hotter than these EUV-emitting post-flare loops. Figure~\ref{fig_OVSA_imaging}a-c show three of the TRACE EUV wavelengths, 171~\AA\ (panel a), 195~\AA\ (panel b), and 284~\AA\ (panel c).  The TRACE images were greatly affected by high-energy particle noise from the earlier X14.4 flare, hence we have processed the images with the eit\_despeckle routine from the Solar Software (SSW) IDL package.  The images show a system of post-flare loops with bright tops centrally located over the active region, with more diffuse loops to both the north and south.  Both footpoints of the post-flare loops appear to be on the disk, just inside the limb.

Hydrogen-alpha images from Big Bear Solar Observatory (BBSO) are available at 16:00 and 17:00 UT.  We show in Figure~\ref{fig_OVSA_imaging}d the relevant portion of the 16:00 UT H$\alpha$ image, showing an earlier stage of post-flare loops.  The BBSO H$\alpha$ image at 17:00 UT shows that the loops with the bright-looptops have migrated southward by then.

\subsection{OVSA Imaging and Relative Visibility}
\label{S_OVSA_Imaging}

As shown in Figure~\ref{fig_OVSA_dynspec}, the OVSA radio burst has significant flux density only in the range 1.2--3.2~GHz.  Figure~\ref{fig_OVSA_imaging} displays a summary of the OVSA imaging at 1.2--3.2~GHz at the peak time of the flare, $\sim$16:42:35~UT, overlaid on the four panels mentioned above. Panels a-c show contours at 20, 40, 60 and 80\% of the peak intensity for 1.2~GHz (panel a), 1.4~GHz (panel b), and 1.8~GHz (panel c). The dashed ellipses in each panel show the sythesized half-power beam size.  Figure~\ref{fig_OVSA_imaging}d shows the relationships of source position and size for the range of frequencies of the burst, as 50\% contours at 1.2~GHz (red), 1.4~GHz (orange), 1.8~GHz (yellow), 2.4~GHz (green) and 3.2~GHz (blue); { see the on-line movie for greater detail}.  The half-power synthesized beam (dashed ellipse) shown in this panel is for the frequency 2.4~GHz.  We have studied the time evolution of the burst at each frequency, and find no significant change of position; { see the on-line movie for greater detail}.  The sources at each frequency rise and fall in intensity from a single location over the 4-5~minute duration of the burst.

The radio sources are spatially associated with the centrally-located post-flare loops--those with bright EUV emission at the loop-tops.  As we will see shortly, the radio spectrum of the burst is consistent with purely thermal gyrosynchrotron emission from a homogeneous source.  However, such a source should have a fixed location and frequency-independent source size, seemingly at odds with the frequency-dependent source sizes and locations shown in Figure~\ref{fig_OVSA_imaging}d.  Of course, much of the source variation could be due to convolution of the source with the frequency-dependent synthesized beam (point spread function).  To check this, we used the simple analytical deconvolution approach of \citet{1970AuJPh..23..113W}, 
which allowed us to use the known 2d-gaussian parameters of the synthesized beam and those of the observed source to derive the parameters of the deconvolved source (semi-major axes $a$ and $b$, and orientation), from which the deconvolved source area $\pi ab$ could be determined. The results of this analysis 
still yield a frequency-dependent \textit{deconvolved} source size ("observed" column of Table~\ref{table}). However, the OVSA interferometric array comprises only 6 antennas, one of which suffered from instrumental effects and was flagged from the data.  Instantaneous (snapshot) imaging with such a small number of antennas (5) provides only a small number of constraints on source parameters, which may be insufficient for an unambiguous determination of deconvolved source size.  To check this possibility, we used the uvmodel routine from the Miriad package \citep{Sault_ea_2011} 
to create a model visibility database using a 2-d gaussian source with fixed source size vs. frequency (we chose the deconvolved 1.8~GHz source area $\pi a b = 4113$~sq. arcsec, with $a = 39.1\arcsec$ and $b = 33.5\arcsec$).  We then sampled the model visibilities with the OVSA antenna spacings.  This model database was then used to create simulated OVSA images, which were deconvolved in the same way as the original OVSA images.  Table~\ref{table} shows the comparison of the original deconvolved source area vs. frequency with that of the simulated images.  It is clear that the same decrease of source area with frequency is seen in both cases, despite the fact that the simulation source model has a constant area.  This implies that determining the source area from the 6-antenna OVRO imaging observations is not reliable, but also that the observations {do not rule out} a roughly constant source size with frequency, as expected for a thermal source.  We will come back to this point in sections~\ref{S_sp_relat} \& \ref{S_Discussion}.

\begin{table}
\caption{ Table 1: Comparison of deconvolved source size for observations and simulation }
\label{table}
\begin{tabular}{cccc}     
  \hline                   
Freq. & Size Observed  & Size Simulated  & $\phi$ Observed \\
(GHz) & (arcsec$^2$) & (arcsec$^2$) & (degrees) \\
  \hline
1.2 & 12742 & 11097 & 346 \\
1.4 & 7241 & 7633 & 343 \\
1.6 & 5252 & 5013 & 345 \\
1.8 & 4113 & 3643 & 341 \\
2.0 & 3552 & 2995 & 329 \\
2.4 & 2301 & 2158 & 348 \\
2.6 & 1736 & 1780 & 355 \\
2.8 & 1205 & 1568 & 350 \\
  \hline
\end{tabular}
\end{table}

Although we have found that the deconvolved source size is not reliable, it is interesting that the orientation of the convolved (observed) sources in Figure~\ref{fig_OVSA_imaging} is slightly different from the orientation of the beam.  The \citet{1970AuJPh..23..113W} 
deconvolution yields an orientation $\phi$ for the deconvolved source of around 345~deg (fourth column of Table~\ref{table}), which is nearly perpendicular to the limb, i.e., in the same direction as the slight dispersion in centroid height with frequency shown in Figure~\ref{fig_OVSA_imaging}d.  This may suggest a vertically-oriented hot loop with a slight frequency asymmetry, i.e. low frequencies tending to come from near the top and high frequencies from near the bottom. However, as we shall see {later}, any such inhomogeneity must be slight enough not to broaden the total power radio spectrum significantly. Stated another way, although the imaging data imply that the sources at different frequencies are displaced relative to each other, they must all have roughly the same effective areas.

{Given the deconvolved source sizes are not reliable, we can study the frequency dependence of the relative visibilities to determine the source size assuming that there is a single source at all frequencies, as in another similar case reported by \citet{Gary_Hurford_1989}.  Recall that the relative visibility is the ratio of baseline amplitude to the square-root of the product of the total power on the two antennas corresponding to that baseline.  It has the advantage that it is independent of calibration.  What one expects to see for a source of constant size \citep[see][]{Gary_Hurford_1989} is a constant \textit{negative slope} vs. frequency$^2$, with the slope proportional to the size of the source.  Contrary to expectations, however, this procedure yields on all baselines a steadily \textit{rising} relative visibility to 3 GHz (points above this are just noise, since the spectrum dies above this).  If taken at face value, this implies a spatially-complex source.  The relative visibility spectral shape is remarkably constant in time throughout the burst, so there is no indication that the source is changing size or shape with time.}

\begin{figure*}\centering
\qquad\includegraphics[width=0.89\textwidth]{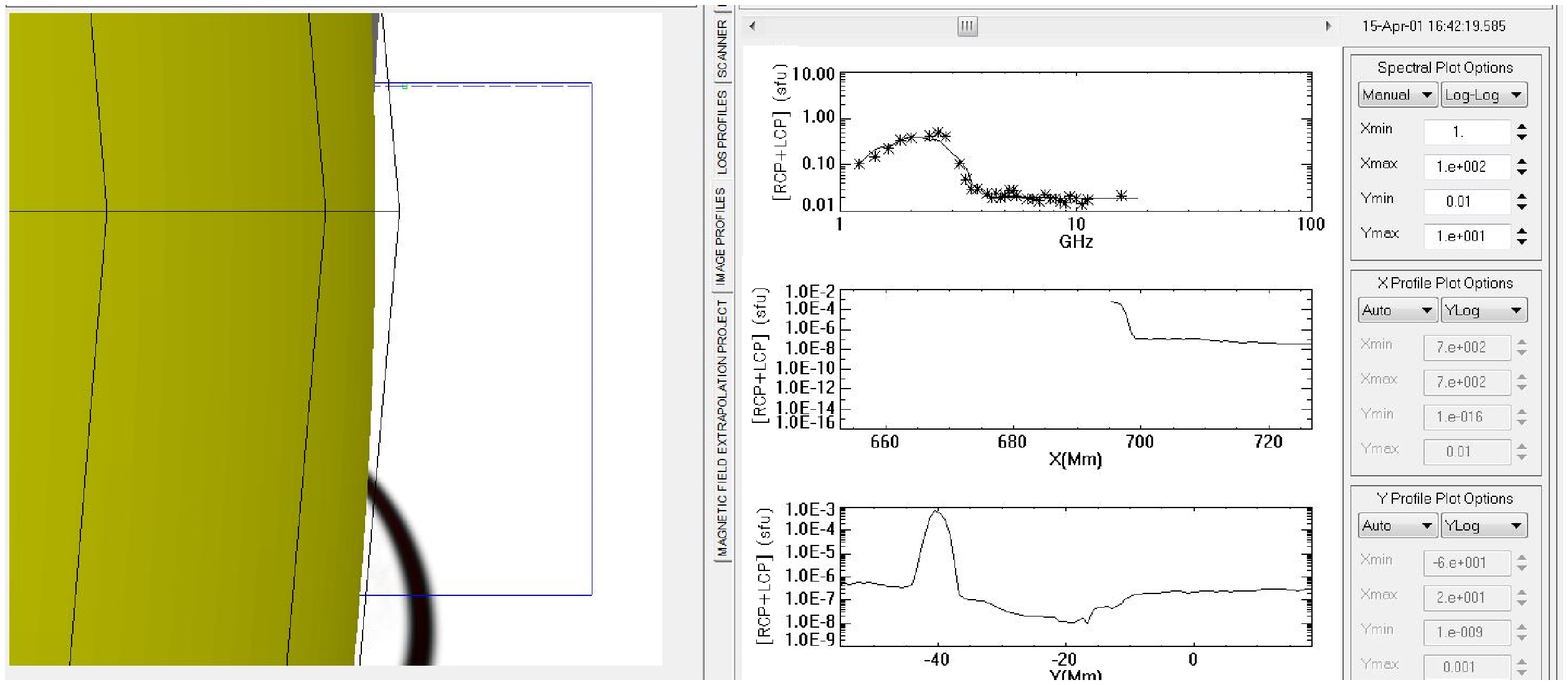}
\includegraphics[width=0.92\textwidth]{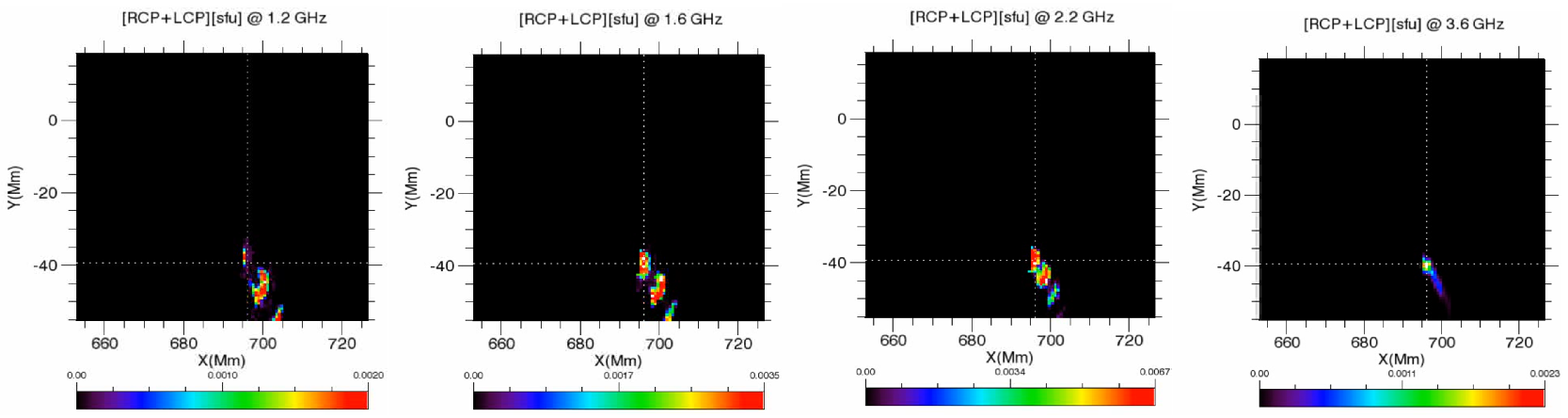}
\caption{\label{fig_model_spectrum_images} A model thermal loop constructed to prove the feasibility of the 'gyro-stripe' model concept. We used a nonlinear force-free extrapolated magnetic data cube obtained for AR 12158 based on the HMI vector magnetogram taken close to the disk center on September 10, 2014. Using the GX Simulator tool \citep{Nita_etal_2015} we selected a magnetic flux tube, rotated the data cube to the western limb, and restricted the field of view to observe the emission from a fraction of the loop seen in the upper left panel. Then, the emission was computed using the 'hybrid' (containing the gyroharmonics) version of the fast code \citep{Fl_Kuzn_2010} to obtain the spectrum (upper right) and the radio images at different frequencies. The set of images (see on-line movie for better impression) proves that the sources at different frequencies can be (i) complex, i.e., consist of a few gyro-stripes embedded in a weaker and more diffuse continuum, (ii) displaced relative to each other, while (iii) have roughly same area and (iv) produce an integrated spectrum (upper right) that nicely fits the observed spectrum, thus, emulating a uniform thermal source. {Note that in a general case the model integrated spectrum will be broader than the observed narrow quasi-uniform thermal spectrum.} 
}
\end{figure*}

\subsection{Understanding the spatial relationships}
\label{S_sp_relat}

{Putting all of the evidence together, we get a picture of a spectral shape and temporal behavior consistent with a relatively compact source of uniform area, but with contrary spatial information that instead indicates a source with significant spatial complexity.  We have insufficient information to be definitive as to how to resolve this quandary, but one possibility is that we are observing the manifestation of discrete gyro-layers expected for non-uniform sources, 
coming from the harmonic spectral features shown in Figure~\ref{Fig_spectrum_example}.  Such features have been suggested before \citep{Kuznetsov_etal_2011}, 
as shown in the movie accompanying the online version of this paper, some frames of which are shown in Figure~\ref{fig_model_spectrum_images}.  Under certain special conditions (which could account for the rarity of such events), the total emitting area of such a non-uniform source, made up of up to 5-7 gyro-layers, could be remarkably constant with frequency over some restricted range, such as the 1.2--3.2~GHz range of the present event, but the centroid of the emission at different frequencies can be displaced following the apparent spatial displacement of the gyro stripes as a function of frequency. Generally, contributions from two different legs of the loop can broaden the spectrum. However, for a loop observed 'edge-on' at the limb---the geometry supported by the recovered source position angle and the source displacement with frequency---the farthest leg can be occulted by the closest one; so only the leg closest to the observer contributes to the spectrum. We return to this point later.
Attempting detailed modeling of this, however, is not well constrained by the data and so beyond the scope of this paper, but the proof-of-concept test in Figure~\ref{fig_model_spectrum_images} suggests that it is feasible.}

\section{Radio Spectral Fit}
\label{S_R_spectral_fit}

\subsection{Fitting framework}
\begin{figure}\centering
\includegraphics[width=0.85\columnwidth]{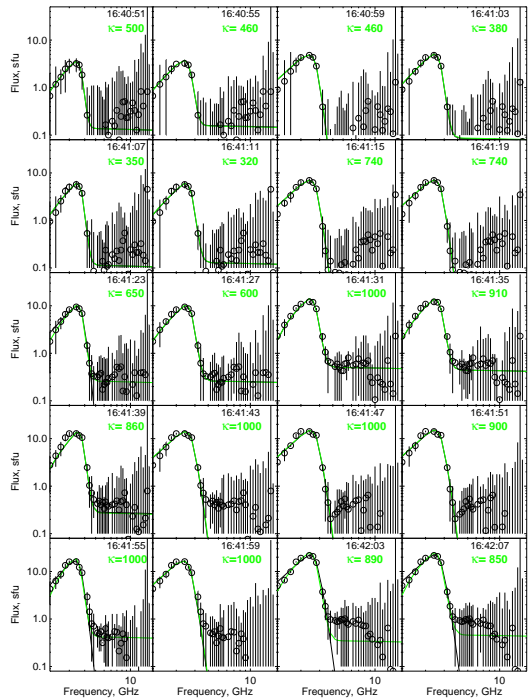}
\caption{\label{fig_fit_example} Examples of the spectrum fits. See the on-line set for the entire flare duration. {Only two significant digits of $\kappa$ are kept.}
}
\end{figure}

In most of the time frames {(see symbols with error bars in Figure~\ref{fig_fit_example})} the microwave spectrum of the burst {rises with frequency $\propto f^2$ at low frequencies until} 
terminated by a very sharp, almost exponential cutoff, which then gives way to a roughly flat flux level at the correspondingly higher frequencies. Such a spectrum
can be fit perfectly by a combination of gyrosynchrotron and free-free emissions from a purely thermal {source with quasi-uniform area as a function of frequency}, where the gyro emission is responsible for the low-frequency, relatively narrowband spectral peak, while the optically thin free-free emission is responsible for the higher-frequency broadband background.

Within this spectral model, the low-frequency region, $\propto f^2$, represents the optically thick thermal gyro emission, thus, the flux level is defined solely by the product of the source area $A$ and the plasma temperature $T$, $F\propto A\cdot T$. In particular, knowing the plasma temperature, e.g., from X-ray data allows estimating the source area \citep{Altyntsev_etal_2012}, while having {information} on the source size allows reading the plasma temperature from the spectrum directly \citep{Gary_Hurford_1989}. There {are} no X-ray data {available} to derive the plasma temperature {for this event}, so we must rely on the {other constraints for} the source size. As has been said in Section~\ref{S_OVSA_Imaging}, the estimate of the source area is not {reliable.} 
It is important, however, that the imaging data do not suggest any noticeable change in the source size or position with time{, while at the same time the narrow spectral shape suggests that the source area cannot vary too much with frequency, as any appreciable variation will broaden the radio spectrum}.
In addition, analysis of the relative visibilities suggests a noticeable spatial complexity of the radio source, which, all together, drove us to a 'gyro-stripe' model envisioned in Sec~\ref{S_sp_relat}, thus, implying the effective source area to comprise only a fraction of the deconvolved source area.
Accordingly, we {loosely} adopt {an area of $A_m = 1600\arcsec^2$, 
which is roughly 40\% of the deconvolved source area found at 1.8~GHz, $A_0 = \pi(39.1\arcsec\times 33.5\arcsec) = 4113\arcsec^2$. One has to keep 
in mind that any temperatures derived from this assumption should be scaled by $A_m/A$, where $A$ is the true source area.  The}
source depth is adopted to be $20\arcsec$, which is an appropriately scaled {deconvolved} source width 
(the exact value of the source depth is {unimportant}; for example, very similar results would be obtained for $10\arcsec$ depth, see Figure~\ref{Fig_spectrum_example}).


A striking feature of this radio burst is nearly perfect constancy of the spectral peak. Indeed, for a thermal source with {constant size,} 
the increase of the flux level can only be provided by {an increase in plasma temperature.}
In a source with a constant magnetic field, this plasma heating must necessarily result in the corresponding increase of the spectrum peak, see Figure~\ref{Fig_spectrum_example}. Thus, the aforementioned constancy of the spectral peak at the radio emitting volume implies {a corresponding} decrease of the magnetic field {over} the course of the plasma heating.

\subsection{Fit to Maxwellian Plasma}
\label{S_M_plasma_fit}

To perform the sequential spectral fit \citep{Bastian_etal_2007, Fl_etal_2009, Fl_etal_2013, Gary_etal_2013} we apply a homogeneous source model with constant depth  $20''$ and area $1600''^2$ and three unknown free parameters  $B(t)$, $T(t)$, and $n_{th}(t)$, the magnetic field, plasma temperature, and density respectively. Selection of the objective function to be forward fit requires some further discussion in this case. Indeed, the gyro emission from a truly homogeneous thermal source (with a unique single magnetic field value), called gyroresonance (GR) emission, would represent a number of distinct narrow peaks occurring around a few small integer multiples of the gyrofrequency (gyroharmonics),  Figure~\ref{Fig_spectrum_example}, although a realistic variation of the magnetic field along the line of sight smooths these peaks out to form a continuum spectrum. Formation of such a continuous spectrum is accounted for by the theory of GR emission, which properly takes into account the magnetic field variation along the line of sight \citep[e.g.,][]{Zheleznyakov_1970} and has been widely applied to the solar active regions \citep{Aliss_1984, Akhmedov_etal_1986, Gary_Hurford_1994, Gary_Hurford_2004, Peterova_etal_2006, Lee_2007, RATAN, Tun_etal_2011, Nita_etal_2011, Kaltman_etal_2012}. For the purpose of the practical forward fit there is no need to include the magnetic field nonuniformity explicitly. Instead, we can use the continuous version of the fast GS code \citep{Fl_Kuzn_2010} from a homogeneous source, which automatically performs the required averaging. We checked via modeling that the results of these two approaches agree very well to each other.

We performed the fit of the observed sequence of the microwave spectra to the continuous fast GS code with the Maxwellian distribution of the radiating electrons using the simplex method with shaking, which proved to work remarkably well for the microwave spectra \citep{Fl_etal_2009, Fl_etal_2013, Gary_etal_2013}. The weights of the individual data points were calculated taking into account the mismatches between different OVSA antenna measurements at the given frequency and time frame. As has been said, there are three free parameters of the model,  $B(t)$, $T(t)$, and $n_{th}(t)$, whose evolution we recovered from the sequential fit. The fit is successful throughout the entire burst duration; examples of the fit are given in Fig.~\ref{fig_fit_example} by black solid curves.

\subsection{Fit to a Plasma with Kappa Distribution}

Interestingly, with this set of the microwave spectra we can address the question of how precisely the Maxwellian distribution is maintained in the radio source. To do so we perform a similar sequential fit but in this case assuming the kappa distribution \citep{vasyliunas68, owocki83, Maksimovic_etal_1997, livadiotis09, pierrard10} of the plasma instead of the Maxwellian one. The kappa distribution {\citep[of the second kind in the notation of][]{livadiotis09}} contains one more free parameter, the index $\kappa$, compared with the Maxwellian distribution, \citep[for the microwave emission, produced by these distributions, see][and references therein]{Chiuderi_Drago_2004, Fl_Kuzn_2010, Fl_Kuzn_2014}. {In a spatially uniform source, t}he value of the index $\kappa$ can be used as a measure of the deviation of the actual plasma distribution from the Maxwellian one in such a way that the kappa distribution with large  $\kappa$ is very similar to the Maxwellian one, while that with small $\kappa$ is markedly different from it. {For example, \citet{2014ApJ...796..142B} have proposed that the kappa distribution can represent a solution of a transport equation for nonthermal electrons in a dense acceleration region, while \citet{2015ApJ...799..129O} found that the X-ray spectra from above-the-loop-top (Masuda) sources can be well fit with the kappa distribution with indices in the range 4--14, although even smaller $\kappa$ indices have been reported by \citet{2009A&A...497L..13K} for the coronal loop-top sources. On the other hand, in a \mw\ total power spectrum, spatial nonuniformity can mimic the effect of finite kappa, i.e. the spectrum of a thermal source will be broadened by the nonuniformity compared with the thermal spectral shape discussed in section~\ref{S_M_plasma_fit}. We checked, via similar modeling to that presented in Fig.~\ref{fig_model_spectrum_images}, that the kappa distribution can often provide an appropriate fit to the total power spectrum from a nonuniform source, even though filled with a purely Maxwellian thermal plasma. Having in mind both of these possibilities, we apply the kappa-distribution fit phenomenologically, without immediate implication for the non-Maxwellian nature of the plasma or the source nonuniformity. }

The fit examples with the kappa distribution are given by the solid green lines in Fig.~\ref{fig_fit_example} along with the $\kappa$ indices obtained for each time frame. Note, that for $\kappa\gtrsim 15-20$ the microwave spectra produced by either Maxwellian or kappa distribution are visually indistinguishable in Fig.~\ref{fig_fit_example} from each other{---the corresponding kappa area is marked yellow in the kappa-index panel of Fig.~\ref{fig_OVSA_fit_parms}}; stated another way the electron energy distribution in the radio source is comparably consistent with both the Maxwellian distribution and the kappa distribution with a reasonably large index over most of the flare duration. However, at the late decay phase the fit with kappa distribution (with progressively smaller $\kappa$ indices) is formally preferable as it yields a significantly  smaller $\chi^2$ measure. Nevertheless, we believe that this does not necessarily indicate a corresponding change of the (local) thermal electron energy distribution, but rather implies a progressively enhanced role of the source nonuniformity due presumably to differing cooling speeds of the adjacent fluxtubes forming the radio source. This further implies that the kappa distribution instead of the Maxwellian one can successfully be used for analysis of the thermal flares with significant spatial nonuniformity.


\subsection{Fit Results}

\begin{figure}\centering
\includegraphics[width=0.75\columnwidth]{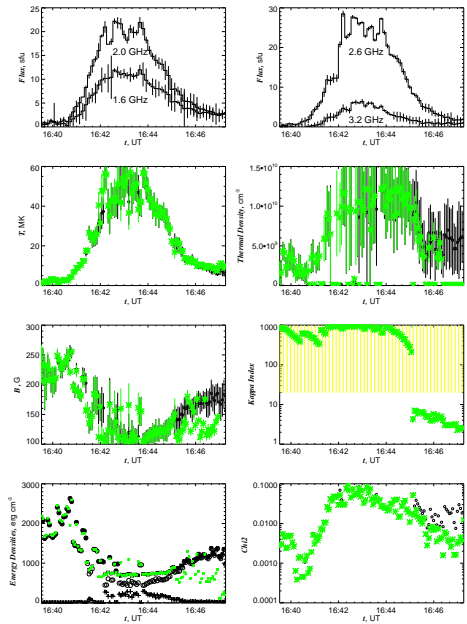}
\caption{\label{fig_OVSA_fit_parms} Radio source parameters  as derived from the OVSA spectral fit  for three parameters of the thermal GS + free-free source.   Two top panels show the radio light curves  at a few frequencies indicated in the plots. The third right panel displays the index of kappa-distribution derived from the fit. {The yellow area indicates the $\kappa$ values that give a distribution indistinguishable from the Maxwellian one; the $\kappa$ uncertainties are of the order of the $\kappa$ values or larger in this area. Below the yellow area the formal uncertainties of the $\kappa$ index are about 10\% and are within the symbol size.} The  lower left panel shows evolution of the energy densities: magnetic (circles) and thermal (plus signs) energy densities as derived from Maxwellian fit and their sum (asterisks) as well as the sum derived from the kappa-distribution fit (green asterisks). The last panel shows the $\chi^2$ evolution for these two fits.
}
\end{figure}

Figure~\ref{fig_OVSA_fit_parms} displays evolution of the fit parameters $B(t)$, $T(t)$, and $n_{th}(t)$ (and the $\kappa$ index in case of kappa distribution) and also the energy densities along with a few radio light curves shown for the reference. The derived plasma temperature, which is almost identical for both Maxwellian and kappa fits, displays clear rise, extended peak, and decay  phases indicative of the plasma heating, saturation, and cooling. The plasma density displays a similar rise-peak-decay pattern although with noticeably larger scatter of the individual data points; the origin of this scatter is apparent: the number density estimate comes from the flat level of the microwave spectrum at high frequencies, where it is measured with large uncertainties. In contrast to these plasma parameters, the magnetic field displays remarkably different behavior: it goes down, stays relatively small, and then increases back to almost original values in case of Maxwellian fit, but in case of the kappa fit it remains at a relatively low level up to the late decay phase.

Interestingly, to obtain a good fit a significant variation of the magnetic field at the source is unavoidable. It is tempting to conclude that we detect here the very process of the magnetic to thermal energy conversion; perhaps, due to the magnetic reconnection process. However, such a straightforward conclusion looks  unwarranted: although the thermal energy density does go up as the magnetic one declines at the flare rise phase, this is insufficient to make up for the magnetic energy density deficit.
Furthermore, the very process of the magnetic to thermal energy conversion would imply the magnetic energy decrease when the plasma temperature keeps roughly constant during the extended peak of the burst to make up for the
significant conductive losses (see next Section); however, the sum of the magnetic and thermal energy keeps remarkably constant there, which requires another energy source to compensate for this loss.

A more likely scenario comes from the loop 'self-occultation' model envisioned in Sec~\ref{S_sp_relat}.
Let us consider a nested loop system, one above the other (with the outer ones having a progressively smaller magnetic field), seen edge-on.
Initially, when the temperature is small (coronal), the outer loops are transparent for the
radio emission at 1.2--2.6 GHz (the observed optically thick part of the spectrum), so we see emission from the most inner loop with the largest
magnetic field ($\sim 200-250$~G). Then the temperature in the outer loops start to grow and they become opaque sequentially---the larger the magnetic field in the given loop the sooner. This way the inner loops become occulted by the outer loop sequentially, so we observe emission from the most outer opaque loop; the hotter the loop the smaller the magnetic field that makes it opaque at the given frequency range; this is
why the fit gives us the proportionally smaller magnetic field for the progressively larger plasma temperature. Apparently, for this model to work we have to adopt that the outer loops are heated progressively stronger than the inner loops. This model is capable of explaining both constancy of the spectral peak during the heating-cooling evolution and constancy of the source location and area in time as observed. Although this 'fine-tuning' of the proposed model may look artificial, we can argue that having a truly exceptional event does require some really seldom combination of physical parameters and/or geometry.



\subsection{A Consistency Check}

The plasma cooling at the burst decay phase must be driven by either radiative or conductive cooling, whose characteristic time scales are strictly specified by the source sizes, temperature, and density defined above. The radiative cooling time \citep[e.g.,][]{Aschw_2005} for such a hot plasma ($T\sim40$~MK) is much longer than the observed one, $\tau\sim30$~s; thus the conductive cooling is likely to dominate. The conductive cooling time, $\tau_{cond}\approx L^2 \rho c_p/(\kappa_S T^{5/2})$ \citep[Eq (4.3.10) in][]{Aschw_2005}, where $L$ is the loop length, $\rho=n_e m_p$ is the plasma mass density, for the  fully ionized hydrogen ideal gas $c_p\rho=5k_B n_e$, where $c_p=2\gamma k_B/[(\gamma-1)m_p]$~erg/(g$\cdot$K) is the specific heat, $\gamma=5/3$, and $\kappa_S\approx 9.2\cdot10^{-7}$~erg cm$^{-1}$ s$^{-1}$ K$^{-7/2}$ is the Spitzer conductivity coefficient \citep[see, e.g.,][]{Aschw_2005, FT_2013}. Solving for $T$ and substituting known constants, we find

\begin{equation}
\label{Eq_T_from_decay}
   T\simeq 3.82\cdot10^7~[{\rm K}] \left(\frac{L}{6\cdot10^9~{\rm cm}}\right)^{4/5}\left(\frac{n_e}{10^{10}~{\rm cm}^{-3}}\right)^{2/5}
   $$$$
   \left(\frac{30~{\rm s}}{\tau}\right)^{2/5},
\end{equation}
which, for the observed source length $L\sim 6\cdot10^9~{\rm cm}$ and decay time scale $\tau\sim30$~s and the plasma density ${n_e}\sim 10^{10}~{\rm cm}^{-3}$ derived from the fit, yields the plasma temperature $T\sim40$~MK in full agreement with the numbers derived from the fit.

\section{Discussion}
\label{S_Discussion}

The ability to precisely derive two components of the flare energy density---the magnetic and thermal ones---in the event under study (where the third generally important energy component, the nonthermal one, is undetectable and so negligible) raises a question of how the energy is being transformed from one form to another and what are the means of the energy dissipation, escape, and build up.

There are two possible scenarios of the event: (1) there is a single (isolated) volume where all these energy transformations occur, in particular, where the magnetic energy is dissipated and then recovers back; or (2) the apparent source of the radio brightness moves in space; in particular, different magnetic field values obtained at different times pertain to physically different spatial locations (layers) rather than reflecting the time evolution of the magnetic field at a given layer.

\subsection{Scenario One: A Single Isolated Source}

In case of an isolated source, there must be a balance between its energy components, two of which are detected. The energy balance can be written in the form
\begin{equation}
\label{Eq_ener_balance}
   \frac{d}{dt} \int\left(\frac{B^2(t)}{8\pi}+\frac{\rho(t) u^2(t)}{2} +3n_{th}(t)k_B T(t) \right)dV=\dot{E}_L,
\end{equation}
where $u(t)$ is the fluid velocity, the integration is performed over the source volume that can change over the course of the flare in a general case, but stays constant (at least, the source area) in our event as suggested by OVSA imaging, and $\dot{E}_L$ is the total energy loss rate, which includes the radiative and conductive losses; the factor of 3 in the third term is written in assumption of the thermal energy equipartition between electrons and ions. It is easy to estimate that both conductive and radiative losses can only remove a  fraction of the thermal energy, which is itself small compared with the total energy dominated by the magnetic field. This means that with a good accuracy we can adopt

\begin{equation}
\label{Eq_ener_cons}
    \int\left(\frac{B^2(t)}{8\pi}+\frac{\rho(t) u^2(t)}{2} +3n_{th}(t)k_B T(t) \right)dV={\rm const}.
\end{equation}
Now, for a source with a constant volume we must require that the sum of the mean energy densities is conserved:
\begin{equation}
\label{Eq_ener_density_cons}
   \frac{B^2(t)}{8\pi}+\frac{\rho(t) u^2(t)}{2} +3n_{th}(t)k_B T(t)=\frac{B_0^2 }{8\pi}
   $$$$
    \approx 2.3\cdot10^3~{\rm erg~cm^{-3}},
\end{equation}
where $B_0$ is the initial magnetic field value at the source. It is easy to estimate that the thermal energy density ($\sim 1.6\cdot10^2$~erg~cm$^{-3}$ at the peak phase) is a minor fraction ($\sim$7\%) of the total energy density and so can be discarded in the zeroth approximation. Then, we note that the magnetic field  at the peak phase of the burst is roughly two times smaller than the initial magnetic field value, see Figure~\ref{fig_OVSA_fit_parms}; thus, the magnetic energy density ($6.7\cdot10^2$~erg~cm$^{-3}$) contains only about 30\% of the original value so the remaining $\sim$70\% must be ascribed to the kinetic energy density, $\rho u^2/2~\sim 1.6\cdot10^3$~erg~cm$^{-3}$, within this model. Such an energy partition cannot be supplied, for example, by an ensemble of MHD waves: in the Alfv\'en wave the kinetic and magnetic energy densities equal to each other; thus, together with the mean magnetic field component, the magnetic energy density must dominate in case of the MHD waves. The dominance of the kinetic energy would imply that the plasma motions are supersonic and super-Alfv\'enic, i.e., the shock waves.
Although not firmly excluded, having such a strong kinetic velocity field in the entire volume seems unlikely for many reasons. One of them is a clear lack of the radio image evolution, which would be expected in the presence of multiple shock waves in the source volume. In addition, such a strong kinetic velocity field would efficiently accelerate the charged particles \citep{Byk_Fl_2009}, which is not observed either.

Finally, there is an option of changing volume (although not supported by observations), when the observed decrease of the magnetic energy density is compensated by the corresponding increase of the flare volume, i.e., the volume expansion. The volume expansion requires a corresponding net driving force. Conversion of the magnetic to thermal energy cannot produce this force since this conversion results in a decrease of the total pressure provided that the total energy is conserved. Thus, the volume expansion necessarily requires a decrease of the magnetic tensions, which can only be achieved through the magnetic reconnection process. This cannot be understood without considering a bigger volume than the immediate radio source.

\subsection{Scenario Two: an Open Source}

If we allow the source of the apparent radio brightness to move along the line of sight (recall, no displacement in the transverse direction is supported by observations) to illuminate  different spatial layers with respectively various magnetic fields detected at the different time frames, we have to consider the energy balance in a bigger volume in which, however, we do not have direct information about spatial distribution of the magnetic field and thermal plasma. Instead, we consider three different stages of the burst---rise, peak, and decay.

During the rise phase we clearly detect the process of the plasma heating during which the hotter the plasma the smaller the magnetic field at the radio source. We have already noted difficulties with a model with magnetic dissipation at a given spatial location to account for this detected decrease of the magnetic field. Thus, we have to conclude that the process of plasma heating moves up from some low layer to higher coronal layers having correspondingly smaller magnetic field as observed. In our 'self-occulting' source model the outer layers are originally cool and, so, optically thin. As soon as the outer layers with smaller magnetic field are heated they become opaque for the gyro process; thus, the lower layers with the stronger magnetic field become invisible. This explains the observed evolution of the magnetic field. Apparently, the heating rate is stronger than the cooling rate at this rise phase.

The extended peak phase of roughly flat light radio curves deserves a close attention. Here, the thermal energy density, although small in the total energy budget, is not small any longer compared with the magnetic energy density at the peak flare phase.  Indeed, a constant temperature level at about 40~MK implies a relatively high heating rate to compensate rapid conductive losses with $\tau_{cond}\sim30$~s. Given that the thermal energy density remains roughly constant at the level of $\sim160$~erg~cm$^{-3}$ over about two minutes, the source volume must have been supplied with $\sim500$~erg~cm$^{-3}$ over those two minutes. This value is comparable to the total magnetic energy density at the radio source during this extended peak phase. Nevertheless, the magnetic field keeps roughly constant during this phase, which unavoidably requires some energy input from an external energy source. Even more remarkable, the sum of the magnetic and thermal energies keeps precisely constant during the peak phase.

When this energy source is switched off or exhausted, the conductive plasma cooling starts to control the further evolution. The higher the temperature the faster the cooling; thus, the effective radiating layer moves back down and the magnetic field increases again almost to the initial level.


\subsection{Energy Distribution of the Flare Plasma}

One more fundamental question that can be conclusively addressed with this particular event, whose spectral evolution is consistent with a purely thermal process of the heating and cooling of the Maxwellian plasma, is what is the accuracy with which the Maxwellian distribution is maintained during the flare, or, equivalently, what deviations from the Maxwellian distribution are consistent with the data. To quantify these deviations we have considered, in Section~\ref{S_R_spectral_fit}, two spectral fits of the microwave spectrum sequence---one with the Maxwellian distribution of the plasma electrons, and the other one with the kappa distribution. A convenience of the kappa distribution for the purpose of characterization of  presumably small deviations of the actual distribution from the Maxwellian one is that it depends on the $\kappa$ index in such a way that the kappa distribution is equivalent to the Maxwellian one for $\kappa\rightarrow\infty$, but develops a more and more pronounced power-law tail for progressively smaller $\kappa$ index. Other parameters of the kappa distribution are the same as for the Maxwellian one; it is important that the temperature $T$ of the kappa distribution has the same physical meaning (the second moment of the distribution characterizing the mean energy of the electrons) as the temperature $T$ of the Maxwellian distribution.

We note that the $\chi^2$ measures\footnote{The $\chi^2$ is noticeably smaller than one because estimates of the the statistical errors are overestimated as they are taken from the small antenna (less sensitive) but applied to big antenna (more sensitive) data. Here we are only interested in relative, rather than absolute, values of the $\chi^2$ measure.} for both fits are comparable to each other over the rise and peak phases of the burst. This implies that the data are comparably well fit by both these distributions. In the case of kappa-distribution fit, the $\kappa$ index is relatively large here, $\kappa\gtrsim 100$, {with typical uncertainties in $\kappa$ larger than 100\%.  Such large values of $\kappa$ result} in visually indistinguishable microwave spectra produced by either Maxwellian or kappa distribution. One can conclude that the data are consistent with the kappa distribution with a large index; but a comparably meaningful conclusion is that there is no objective evidence in favor of any deviation from the Maxwellian distribution over the rise and peak phases. Importantly, the recovered physical parameters of the source, $B(t)$, $T(t)$, and $n_{th}(t)$, are all consistent for both fits over these phases.

The decay phase is apparently different from the rise and peak phases. Indeed, the $\chi^2$ is clearly smaller for the kappa distribution than for the Maxwellian one, which formally favors the kappa over the Maxwellian distribution. The $\kappa$ index decreases progressively at the decay phase from 8 to 2, {with the uncertainty about 10\%,} the magnetic field derived from the kappa fit is noticeably smaller than that derived from the Maxwellian fit; however, the temperatures still agree very well with each other.

Thus, the microwave spectra at the decay phase are apparently better consistent with the kappa than Maxwellian fit, which does not allow to firmly conclude that the Maxwellian distribution is maintained over the decay phase. One possible reason is that the actual energy distribution does deviate from the Maxwellian one at the decay phase due, perhaps, to nonthermal electron generation. It does not sound plausible, however, because this would imply that the particle acceleration did not take place at the stage of essential energy release, but appeared only at the decay phase, after the energy release is over. {Instead, we believe it more likely} that the plasma electron distribution remains (more or less) Maxwellian even at the decay phase; however, a nonuniform cooling results in a multi-temperature instead of the single-temperature plasma volume, so the deviation from the Maxwellian fit at the decay phase is indicative of increasing source nonuniformity, rather than a nonthermal electron generation. This further implies that a uniform source with the kappa distribution can be used as a working model for  nonuniform multi-temperature Maxwellian sources in other, more complex thermal solar flares.

\section{Conclusions}

We have described a nice solar flare whose spectral evolution in the microwave domain is consistent with a purely thermal behavior; only very minor deviations (if any) from the Maxwellian distribution are consistent with the data. From this perspective this event is similar to a thermal flare reported earlier by \citet{Gary_Hurford_1989}, whose spatial structures are, however, strikingly different from each other (a single source in the former case, while a complex source in our case). We believe that this difference comes from the different flare locations in these two cases. \citet{Gary_Hurford_1989} reported a flare seen on disk, where the source of the gyro emission is observed at a quasi-parallel direction to the magnetic field, where no gyro-stripe structure is present. On the contrary, our event happened at the limb, so we likely observe a loop quasi-transversely to the magnetic field, when the spectral harmonic structure and the spatial gyro-stipes are the most pronounced.

We detected a clear evolution of the source parameters such as $B(t)$, $T(t)$, and $n_{th}(t)$ suggesting a corresponding evolution of the magnetic and thermal energies at the radio source. However, the measured energy components do not easily obey the energy conservation requirement, implying that the observed source does not represent an isolated volume, but rather efficiently exchanges its energy with a bigger surrounding volume. In particular, we concluded that the apparent decrease of the magnetic field at the radio source over the rise phase of the flare cannot be easily associated with the magnetic energy transformation to the thermal or kinetic energy within a single isolated volume, but instead requires  an upward propagating magnetic reconnection/plasma heating process, such as in the standard flare scenario with one remarkable difference, however: the absence of the nonthermal electron generation. We hope that the study of other thermal flares will better clarify the origin of such purely thermal events, which show a significant energy release observed through the plasma heating, but do not result in any measurable acceleration of the charged particles.

\acknowledgments
This work was supported in part by NSF grants  AGS-1250374 and AGS-1262772 and NASA grants NNX10AF27G and NNX14AC87G to New Jersey Institute of Technology.
This work also benefited from workshop support from the International Space Science Institute (ISSI).

\newpage

\bibliographystyle{apj}
\bibliography{xray_refs,fleishman,ms_Thermal_Flare,AR_bib}

\begin{thebibliography}{51}
\expandafter\ifx\csname natexlab\endcsname\relax\def\natexlab#1{#1}\fi

\bibitem[{{Akhmedov} {et~al.}(1986){Akhmedov}, {Borovik}, {Gelfreikh}, {Bogod},
  {Korzhavin}, {Petrov}, {Dikij}, {Lang}, \& {Willson}}]{Akhmedov_etal_1986}
{Akhmedov}, S.~B., {et~al.} 1986, \apj, 301, 460

\bibitem[{{Alissandrakis} \& {Kundu}(1984)}]{Aliss_1984}
{Alissandrakis}, C.~E., \& {Kundu}, M.~R. 1984, \aap, 139, 271

\bibitem[{{Altyntsev} {et~al.}(2012){Altyntsev}, {Fleishman}, {Lesovoi}, \&
  {Meshalkina}}]{Altyntsev_etal_2012}
{Altyntsev}, A.~A., {Fleishman}, G.~D., {Lesovoi}, S.~V., \& {Meshalkina},
  N.~S. 2012, \apj, 758, 138

\bibitem[{{Asai} {et~al.}(2006){Asai}, {Nakajima}, {Shimojo}, {White},
  {Hudson}, \& {Lin}}]{Asai_etal_2006}
{Asai}, A., {Nakajima}, H., {Shimojo}, M., {White}, S.~M., {Hudson}, H.~S., \&
  {Lin}, R.~P. 2006, \pasj, 58, L1

\bibitem[{{Asai} {et~al.}(2009){Asai}, {Nakajima}, {Shimojo}, {Yokoyama},
  {Masuda}, \& {Krucker}}]{Asai_etal_2009}
{Asai}, A., {Nakajima}, H., {Shimojo}, M., {Yokoyama}, T., {Masuda}, S., \&
  {Krucker}, S. 2009, \apj, 695, 1623

\bibitem[{{Aschwanden}(2005)}]{Aschw_2005}
{Aschwanden}, M.~J. 2005, {Physics of the Solar Corona. An Introduction with
  Problems and Solutions (2nd edition)} (Pour la Science)

\bibitem[{{Bastian} {et~al.}(2007){Bastian}, {Fleishman}, \&
  {Gary}}]{Bastian_etal_2007}
{Bastian}, T.~S., {Fleishman}, G.~D., \& {Gary}, D.~E. 2007, \apj, 666, 1256

\bibitem[{{Battaglia} {et~al.}(2009){Battaglia}, {Fletcher}, \&
  {Benz}}]{Battaglia_etal_2009}
{Battaglia}, M., {Fletcher}, L., \& {Benz}, A.~O. 2009, \aap, 498, 891

\bibitem[{{Benka} \& {Holman}(1992)}]{Benka_Holman_1992}
{Benka}, S.~G., \& {Holman}, G.~D. 1992, \apj, 391, 854

\bibitem[{{Bian} {et~al.}(2014){Bian}, {Emslie}, {Stackhouse}, \&
  {Kontar}}]{2014ApJ...796..142B}
{Bian}, N.~H., {Emslie}, A.~G., {Stackhouse}, D.~J., \& {Kontar}, E.~P. 2014,
  \apj, 796, 142

\bibitem[{{Bogod} {et~al.}(2000){Bogod}, {Garaimov}, {Zheleznyakov}, \&
  {Zlotnik}}]{Bogod_etal_2000}
{Bogod}, V.~M., {Garaimov}, V.~I., {Zheleznyakov}, V.~V., \& {Zlotnik}, E.~Y.
  2000, Astronomy Reports, 44, 271

\bibitem[{{Bogod} \& {Yasnov}(2009)}]{RATAN}
{Bogod}, V.~M., \& {Yasnov}, L.~V. 2009, Astrophysical Bulletin, 64, 372

\bibitem[{{Bykov} \& {Fleishman}(2009)}]{Byk_Fl_2009}
{Bykov}, A.~M., \& {Fleishman}, G.~D. 2009, \apjl, 692, L45

\bibitem[{{Chiuderi} \& {Chiuderi Drago}(2004)}]{Chiuderi_Drago_2004}
{Chiuderi}, C., \& {Chiuderi Drago}, F. 2004, \aap, 422, 331

\bibitem[{{Delaboudini{\`e}re} {et~al.}(1995){Delaboudini{\`e}re}, {Artzner},
  {Brunaud}, {Gabriel}, {Hochedez}, {Millier}, {Song}, {Au}, {Dere}, {Howard},
  {Kreplin}, {Michels}, {Moses}, {Defise}, {Jamar}, {Rochus}, {Chauvineau},
  {Marioge}, {Catura}, {Lemen}, {Shing}, {Stern}, {Gurman}, {Neupert},
  {Maucherat}, {Clette}, {Cugnon}, \& {van Dessel}}]{soho_eit}
{Delaboudini{\`e}re}, J.-P., {et~al.} 1995, \solphys, 162, 291

\bibitem[{{Domingo} {et~al.}(1995){Domingo}, {Fleck}, \& {Poland}}]{soho}
{Domingo}, V., {Fleck}, B., \& {Poland}, A.~I. 1995, \solphys, 162, 1

\bibitem[{{Emslie} {et~al.}(2012){Emslie}, {Dennis}, {Shih}, {Chamberlin},
  {Mewaldt}, {Moore}, {Share}, {Vourlidas}, \& {Welsch}}]{Emslie_etal_2012}
{Emslie}, A.~G., {et~al.} 2012, \apj, 759, 71

\bibitem[{{Fleishman} {et~al.}(2011){Fleishman}, {Kontar}, {Nita}, \&
  {Gary}}]{Fl_etal_2011}
{Fleishman}, G.~D., {Kontar}, E.~P., {Nita}, G.~M., \& {Gary}, D.~E. 2011,
  \apjl, 731, L19

\bibitem[{{Fleishman} {et~al.}(2013){Fleishman}, {Kontar}, {Nita}, \&
  {Gary}}]{Fl_etal_2013}
---. 2013, \apj, 768, 190

\bibitem[{{Fleishman} \& {Kuznetsov}(2010)}]{Fl_Kuzn_2010}
{Fleishman}, G.~D., \& {Kuznetsov}, A.~A. 2010, \apj, 721, 1127

\bibitem[{{Fleishman} \& {Kuznetsov}(2014)}]{Fl_Kuzn_2014}
---. 2014, \apj, 781, 77

\bibitem[{{Fleishman} {et~al.}(2009){Fleishman}, {Nita}, \&
  {Gary}}]{Fl_etal_2009}
{Fleishman}, G.~D., {Nita}, G.~M., \& {Gary}, D.~E. 2009, \apjl, 698, L183

\bibitem[{{Fleishman} \& {Toptygin}(2013)}]{FT_2013}
{Fleishman}, G.~D., \& {Toptygin}, I.~N. 2013, Cosmic Electrodynamics.
  Astrophysics and Space Science Library; Springer NY, Vol. 388, {712 p [FT13]}

\bibitem[{{Gary} {et~al.}(2013){Gary}, {Fleishman}, \& {Nita}}]{Gary_etal_2013}
{Gary}, D.~E., {Fleishman}, G.~D., \& {Nita}, G.~M. 2013, \solphys, 288, 549

\bibitem[{{Gary} \& {Hurford}(1989)}]{Gary_Hurford_1989}
{Gary}, D.~E., \& {Hurford}, G.~J. 1989, \apj, 339, 1115

\bibitem[{{Gary} \& {Hurford}(1994)}]{Gary_Hurford_1994}
---. 1994, \apj, 420, 903

\bibitem[{{Gary} \& {Hurford}(2004)}]{Gary_Hurford_2004}
{Gary}, D.~E., \& {Hurford}, G.~J. 2004, in Astrophysics and Space Science
  Library, Vol. 314, Astrophysics and Space Science Library, ed. D.~E. {Gary}
  \& C.~U. {Keller}, 71

\bibitem[{{Handy} {et~al.}(1999){Handy}, {Acton}, {Kankelborg}, {Wolfson},
  {Akin}, {Bruner}, {Caravalho}, {Catura}, {Chevalier}, {Duncan}, {Edwards},
  {Feinstein}, {Freeland}, {Friedlaender}, {Hoffmann}, {Hurlburt}, {Jurcevich},
  {Katz}, {Kelly}, {Lemen}, {Levay}, {Lindgren}, {Mathur}, {Meyer}, {Morrison},
  {Morrison}, {Nightingale}, {Pope}, {Rehse}, {Schrijver}, {Shine}, {Shing},
  {Strong}, {Tarbell}, {Title}, {Torgerson}, {Golub}, {Bookbinder}, {Caldwell},
  {Cheimets}, {Davis}, {Deluca}, {McMullen}, {Warren}, {Amato}, {Fisher},
  {Maldonado}, \& {Parkinson}}]{trace}
{Handy}, B.~N., {et~al.} 1999, \solphys, 187, 229

\bibitem[{{Hurford} {et~al.}(1984){Hurford}, {Read}, \& {Zirin}}]{ovsa_1984}
{Hurford}, G.~J., {Read}, R.~B., \& {Zirin}, H. 1984, \solphys, 94, 413

\bibitem[{{Ji{\v r}i{\v c}ka} {et~al.}(2001){Ji{\v r}i{\v c}ka},
  {Karlick{\'y}}, {M{\'e}sz{\'a}rosov{\'a}}, \& {Sn{\'{\i}}{\v
  z}ek}}]{ondrejov}
{Ji{\v r}i{\v c}ka}, K., {Karlick{\'y}}, M., {M{\'e}sz{\'a}rosov{\'a}}, H., \&
  {Sn{\'{\i}}{\v z}ek}, V. 2001, \aap, 375, 243

\bibitem[{{Kaltman} {et~al.}(2012){Kaltman}, {Bogod}, {Stupishin}, \&
  {Yasnov}}]{Kaltman_etal_2012}
{Kaltman}, T.~I., {Bogod}, V.~M., {Stupishin}, A.~G., \& {Yasnov}, L.~V. 2012,
  Astronomy Reports, 56, 790

\bibitem[{{Ka{\v s}parov{\'a}} \& {Karlick{\'y}}(2009)}]{2009A&A...497L..13K}
{Ka{\v s}parov{\'a}}, J., \& {Karlick{\'y}}, M. 2009, \aap, 497, L13

\bibitem[{{Kuznetsov} {et~al.}(2011){Kuznetsov}, {Nita}, \&
  {Fleishman}}]{Kuznetsov_etal_2011}
{Kuznetsov}, A.~A., {Nita}, G.~M., \& {Fleishman}, G.~D. 2011, ArXiv e-prints

\bibitem[{{Lang} {et~al.}(1987){Lang}, {Willson}, {Smith}, \&
  {Strong}}]{Lang_etal_1987}
{Lang}, K.~R., {Willson}, R.~F., {Smith}, K.~L., \& {Strong}, K.~T. 1987, \apj,
  322, 1044

\bibitem[{{Lee}(2007)}]{Lee_2007}
{Lee}, J. 2007, \ssr, 133, 73

\bibitem[{{Liu} {et~al.}(2013){Liu}, {Li}, \& {Fletcher}}]{liu_etal_2013}
{Liu}, S., {Li}, Y., \& {Fletcher}, L. 2013, \apj, 769, 135

\bibitem[{{Livadiotis} \& {McComas}(2009)}]{livadiotis09}
{Livadiotis}, G., \& {McComas}, D.~J. 2009, Journal of Geophysical Research
  (Space Physics), 114, 11105

\bibitem[{{Maksimovic} {et~al.}(1997){Maksimovic}, {Pierrard}, \&
  {Lemaire}}]{Maksimovic_etal_1997}
{Maksimovic}, M., {Pierrard}, V., \& {Lemaire}, J.~F. 1997, \aap, 324, 725

\bibitem[{{Neupert}(1968)}]{1968ApJ...153L..59N}
{Neupert}, W.~M. 1968, \apjl, 153, L59

\bibitem[{{Nita} {et~al.}(2011){Nita}, {Fleishman}, {Jing}, {Lesovoi}, {Bogod},
  {Yasnov}, {Wang}, \& {Gary}}]{Nita_etal_2011}
{Nita}, G.~M., {Fleishman}, G.~D., {Jing}, J., {Lesovoi}, S.~V., {Bogod},
  V.~M., {Yasnov}, L.~V., {Wang}, H., \& {Gary}, D.~E. 2011, \apj, 737, 82

\bibitem[{{Nita} {et~al.}(2014){Nita}, {Fleishman}, {Kuznetsov}, {Kontar}, \&
  {Gary}}]{Nita_etal_2015}
{Nita}, G.~M., {Fleishman}, G.~D., {Kuznetsov}, A.~A., {Kontar}, E.~P., \&
  {Gary}, D.~E. 2014, ArXiv e-prints

\bibitem[{{Nita} {et~al.}(2004){Nita}, {Gary}, \& {Lee}}]{Nita_etal_2004}
{Nita}, G.~M., {Gary}, D.~E., \& {Lee}, J. 2004, \apj, 605, 528

\bibitem[{{Oka} {et~al.}(2015){Oka}, {Krucker}, {Hudson}, \&
  {Saint-Hilaire}}]{2015ApJ...799..129O}
{Oka}, M., {Krucker}, S., {Hudson}, H.~S., \& {Saint-Hilaire}, P. 2015, \apj,
  799, 129

\bibitem[{{Owocki} \& {Scudder}(1983)}]{owocki83}
{Owocki}, S.~P., \& {Scudder}, J.~D. 1983, \apj, 270, 758

\bibitem[{{Peterova} {et~al.}(2006){Peterova}, {Agalakov}, {Borisevich},
  {Korzhavin}, \& {Ryabov}}]{Peterova_etal_2006}
{Peterova}, N.~G., {Agalakov}, B.~V., {Borisevich}, T.~P., {Korzhavin}, A.~N.,
  \& {Ryabov}, B.~I. 2006, Astronomy Reports, 50, 679

\bibitem[{{Pierrard} \& {Lazar}(2010)}]{pierrard10}
{Pierrard}, V., \& {Lazar}, M. 2010, \solphys, 267, 153

\bibitem[{{Sault} {et~al.}(2011){Sault}, {Teuben}, \& {Wright}}]{Sault_ea_2011}
{Sault}, R.~J., {Teuben}, P.~J., \& {Wright}, M.~C.~H. 2011, Astrophysics
  Source Code Library, 6007

\bibitem[{{Tun} {et~al.}(2011){Tun}, {Gary}, \& {Georgoulis}}]{Tun_etal_2011}
{Tun}, S.~D., {Gary}, D.~E., \& {Georgoulis}, M.~K. 2011, \apj, 728, 1

\bibitem[{{Vasyliunas}(1968)}]{vasyliunas68}
{Vasyliunas}, V.~M. 1968, in Astrophysics and Space Science Library, Vol.~10,
  Physics of the Magnetosphere, ed. R.~D.~L. {Carovillano} \& J.~F. {McClay},
  622

\bibitem[{{Wild}(1970)}]{1970AuJPh..23..113W}
{Wild}, J.~P. 1970, Australian Journal of Physics, 23, 113

\bibitem[{{Zheleznyakov}(1970)}]{Zheleznyakov_1970}
{Zheleznyakov}, V.~V. 1970, {Radio emission of the sun and planets}

\end{thebibliography}





\end{document}